\documentclass[journal]{IEEEtran}
\usepackage[utf8]{inputenc}
\usepackage{float}
 
\usepackage[ruled, linesnumbered]{algorithm2e} %
\usepackage{amsfonts}
\usepackage{amsmath}
\usepackage{amsthm}
\usepackage{amssymb}
\usepackage{array}
\usepackage{bm}
\usepackage{cases}
\usepackage{color}
\usepackage[numbers,sort&compress]{natbib}
\usepackage[font=footnotesize,labelfont=bf]{caption}
\usepackage{enumitem}

\usepackage[overload]{empheq}
\usepackage{graphicx}   %use graph format
\usepackage{makecell}
\usepackage{mathrsfs}
\usepackage{multirow} 
\usepackage{comment}
\usepackage{verbatim}

\usepackage{physics}    % for \dv
\usepackage{soul,color}
\usepackage{subcaption}
\usepackage{url}
\usepackage[all]{xy}

\usepackage{threeparttable}
\usepackage{tabularx}
\usepackage{longtable}
\usepackage{xcolor}

\usepackage{lineno}

\usepackage{xspace}
\usepackage{float}

\usepackage{blindtext}
\usepackage{hyperref}
\usepackage{nameref}

\usepackage{tikz}    
\usepackage{standalone}
\usetikzlibrary{arrows , arrows.meta, calc , fit}
\usetikzlibrary{shapes, chains, fit,backgrounds,shapes,positioning}
\usepackage{marvosym}

\makeatletter
\def\subsubsection{\@startsection{subsubsection}% name
                                 {3}% level
                                 {\z@}% indent (formerly \parindent)
                                 {0ex plus 0.1ex minus 0.1ex}% before skip
                                 {0ex}% after skip
                                 {\normalfont\normalsize\itshape}}% style
\makeatother

\newcommand{\PreserveBackslash}[1]{\let\temp=\\#1\let\\=\temp}
\newcolumntype{C}[1]{>{\PreserveBackslash\centering}p{#1}}
\newcolumntype{R}[1]{>{\PreserveBackslash\raggedleft}p{#1}}
\newcolumntype{L}[1]{>{\PreserveBackslash\raggedright}p{#1}}

\begin{document}

\title{  

Resilient Energy-Based Control for DC Data Centers under Grid and Load Disturbances}
\author{Lizhi~Wang,~\IEEEmembership{Member,~IEEE},
Fei~Feng,~\IEEEmembership{Member,~IEEE},  Ella~Chou,~\IEEEmembership{Member,~IEEE}, and
Yashen~Lin, ~\IEEEmembership{Member,~IEEE}
\thanks{L. Wang, E. Chou and Y. Lin are with Siemens Foundational Technology, Princeton, NJ, USA (e-mail: wang.lizhi@siemens.com).}
\thanks{F. Feng  is with the Department of Electrical Engineering, State University of New York, Maritime College, Bronx, NY 10465, USA } 
}
\markboth{}
{Shell \MakeLowercase{\textit{ et al.}}:  Bare Demo of IEEEtran.cls for IEEE Journals}

\maketitle
\begin{abstract}
This paper presents a passivity-based control framework for AC/DC converters supplying non-passive Information Technology(IT) rack loads in DC data centers. Unlike conventional cascaded proportional–integral (PI) controllers that ensure stability only near nominal operating points, the proposed method is derived from the system’s total energy balance using the Port-Hamiltonian (PH) formulation. By shaping the stored energy and injecting virtual damping through a lossless interconnection with a PH controller, the converter behaves as a passive system even when interfaced with non-passive loads or under grid disturbances. The closed-loop system guarantees asymptotic voltage regulation and strict energy dissipation without assuming constant grid voltage or frequency. Simulation studies under realistic load and fault scenarios validate that the proposed controller achieves smaller voltage deviations, faster recovery, and superior robustness, demonstrating its suitability for future high-efficiency DC data-center architectures.
\end{abstract}

\begin{IEEEkeywords}
DC data center, Port-Hamiltonian, AC/DC converter, Large signal stability
\end{IEEEkeywords}
 
 \section{Introduction}
\IEEEPARstart{T}{he} rapid growth of artificial-intelligence and high-performance-computing workloads has driven data-center power densities beyond hundreds of kilowatts per rack~\cite{davenport2024ai}. Traditional AC distribution with double-conversion UPS architectures suffers from repeated AC/DC conversions, limited efficiency, and complex fault coordination. DC architectures have therefore emerged as a promising alternative, reducing conversion stages and enabling integration with renewable and storage systems~\cite{OCP_Diablo400_2025}. However, DC data centers~\cite{krein2017data} introduce new dynamic-stability challenges, primarily due to tightly regulated downstream converters that exhibit constant-power-load (CPL) characteristics and inject negative incremental resistance into the DC bus~\cite{beheshtaein2019dc}. Moreover, when the grid experiences voltage or frequency deviations, data centers are often required to disconnect rapidly to protect critical IT equipment. Such islanding or large-load shedding events, however, can cause severe grid disturbances and propagate cascading impacts across the system~\cite{NERC_LargeLoads_2025}.

Traditional double-loop PI controllers~\cite{zhang2021networked} regulation are typically tuned empirically and ensure only small-signal stability around nominal operating points. Their design relies on linearized models that neglect the strong nonlinear coupling between the AC and DC sides of the converter as well as the negative incremental impedance introduced by constant-power loads. As a result, PI-controlled rectifiers behave as fixed admittance systems whose stability margin deteriorates rapidly when the load power, grid impedance, or DC-bus voltage deviate from the tuning condition. The inner current loop, designed to enforce fast current tracking, interacts with the outer voltage loop through the DC-link dynamics, creating multiple feedback paths that are highly sensitive to parameter variation. Under large disturbances—such as sudden load steps, voltage sags, or frequency deviations—the linear PI gains cannot adapt to the changing energy balance, causing oscillatory or divergent responses~\cite{gui2021large}. The PH-based control has been investigated to ensure large signal stability in power-electronic systems but with differing scopes. A PH framework~\cite{zhong2021port} is designed to render converters passive via a lossless interconnection. PH concept are extended to a modified dispatchable virtual-oscillator control (dVOC) for grid-forming inverters~\cite{kong2023control}, achieving passivity and transient stability without assuming constant voltage or frequency, yet focusing mainly on AC synchronization rather than DC interface dynamics. Despite its theoretical strength, PH control has not been fully explored for grid-connected DC architectures with non-passive data-center loads.

To bridge this gap,  this paper develops a PH-based passivity-rendering controller for grid-tied AC/DC converters supplying non-passive loads in DC data centers. The grid, converter, and DC-link subsystems are modeled consistently within the PH framework to preserve their physical energy interconnections. The control law ensures that the total stored energy monotonically decreases in the absence of external excitation by combining an outer energy-shaping voltage loop and an inner damping-injection current loop. This passivity-rendering mechanism achieves stable operation under CPLs and dynamic grid conditions, maintaining DC-bus voltage regulation, enabling dissipative grid interaction, and preventing large-load shedding during transients.

The remainder of this paper is organized as follows. Section~II introduces the PH framework and system modeling. Section~III details the passivity-based control law. Section~IV presents simulation case studies validating the proposed design, and Section~V concludes the paper.

\vspace{-10pt}
\section{System Modeling}
\subsection{Port-Hamiltonian Framework}

The PH framework provides a unified method for describing multi-domain energy systems using energy functions and power-preserving interconnections. 
Unlike conventional state-space models, PH representations express the dynamics directly in terms of stored energy and power exchange, 
making the physical meaning of each variable explicit.
A general PH system can be written as\cite{zhong2021port}
\begin{equation}
\dot{x} = (\mathcal{J} - \mathcal{R}) \nabla H(x) + G\,u, 
\qquad 
y = G^\top \nabla H(x),
\label{eq:ph_general}
\end{equation}
where
\begin{itemize}
    \item $x \in \mathbb{R}^n$ is the vector of \emph{energy variables} (e.g., flux linkages, charges, or momenta);
    \item $H(x):\mathbb{R}^n \to \mathbb{R}_+$ is the \emph{Hamiltonian}, representing the total stored energy;
    \item $\nabla H(x)$ is the \emph{co-energy} vector (generalized efforts);
    \item $\mathcal{J} = -\mathcal{J}^\top$ is the \emph{interconnection matrix}, encoding lossless power exchange between storage elements;
    \item $\mathcal{R} = \mathcal{R}^\top \succeq 0$ is the \emph{dissipation matrix}, modeling resistive or damping losses;
    \item $u$ and $y$ are the \emph{port variables} (effort and flow) satisfying the power relation $u^\top y$.
\end{itemize}

Taking the time derivative of the Hamiltonian yields the fundamental PH energy balance:
\begin{equation}
\dot{H} = (\nabla H)^\top \dot{x} 
         = u^\top y - (\nabla H)^\top \mathcal{R} \nabla H.
\label{eq:ph_energy_balance}
\end{equation}
Thus, the internal energy increases only through external power input and decreases through dissipative losses. 
If $\mathcal{R}=0$, the system is lossless; if $\mathcal{R}\succeq0$, it is passive from input $u$ to output $y$.

Multiple PH subsystems can be interconnected through \emph{power-conserving} Dirac structures that preserve the overall energy balance. When the port efforts and flows satisfy $u_1^{\top}y_1 + u_2^{\top}y_2 = 0$, the composite system remains PH with the total Hamiltonian $H_{\text{total}} = \sum_i H_i$. Because PH systems are inherently passive, they can be stabilized through \emph{control by interconnection}, where the controller is itself a PH subsystem characterized by its own energy function and dissipation matrix. When the plant and controller are interconnected through a lossless port, the closed-loop system retains the PH structure, and stability is achieved by shaping the combined Hamiltonian to make the desired equilibrium a minimum while injecting additional damping to ensure asymptotic convergence.
\vspace{-10pt}
\subsection{DC data center power architecture in Port-Hamiltonian Framework}
A simplified $L$--$C_{dc}$ model is adopted, neglecting the AC-side shunt capacitor $C_f$, which is acceptable since the converter operates as a rectifier (grid-following mode) and does not need to synthesize a sinusoidal voltage.
\vspace{-10pt}
\begin{figure}[ht]
\centering
\includegraphics[width=0.5\textwidth]{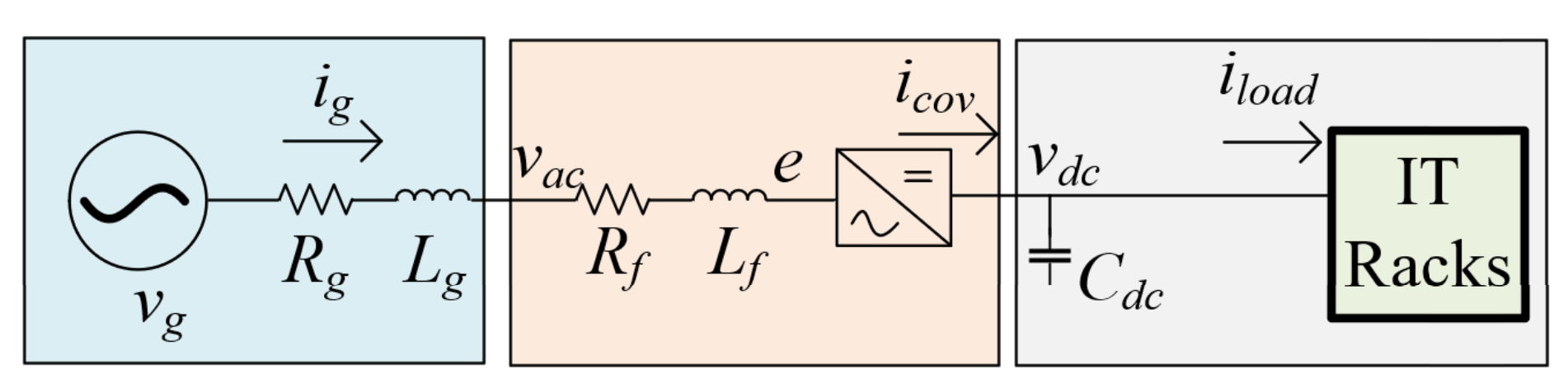}
\caption{DC data center power architecture. } 
\label{ac/dc}
\end{figure} 
\vspace{-10pt}
As illustrated in Fig.~\ref{ac/dc}, the complete network is partitioned into three physically meaningful subsystems:
\begin{itemize}
    \item \textbf{Subsystem $\boldsymbol{\Sigma_{grid}}$ -- Grid and Line:} the external AC source with impedance $(R_g,\,L_g)$;
    \item \textbf{Subsystem $\boldsymbol{\Sigma_{cov}}$ -- Converter and Filter:} the controlled AC/DC interface including RL filter $(R_f,\,L_f)$ and the controlled rectifier $e$;
    \item \textbf{Subsystem $\boldsymbol{\Sigma_{dc}}$ -- DC-Link and Load:} the DC capacitor $C_{dc}$ and the downstream load branch representing the aggregated IT-rack power converters.
\end{itemize}

Each subsystem is modeled individually in the Port-Hamiltonian (PH) form and then interconnected through power-preserving Dirac structures to yield the global energy-consistent model.
\vspace{-10pt}
\subsection{Grid and Line Subsystem $\Sigma_{\text{grid}}$}
The grid is modeled as a balanced voltage source $v_g=[v_{g\alpha},v_{g\beta}]^\top$ behind an impedance $R_g+sL_g$.
The inductor flux $\phi_g=L_g i_g$ is selected as the energy state, and the stored energy is
\begin{equation}
H_g(\phi_g)=\tfrac{1}{2L_g}\|\phi_g\|^2 .
\end{equation}
Using $i_g=\nabla_{\phi_g}H_g=\phi_g/L_g$, the PH formulation of the grid subsystem is
\begin{equation}
\Sigma_{\text{grid}}:
\begin{cases}
\dot{\phi}_g=(\mathcal J_g-\mathcal R_g)\nabla H_g + G_g v_g,\\[2pt]
y_g=G_g^{\!\top}\nabla H_g=i_g,
\end{cases}
\end{equation}
where $\mathcal J_g=0$, $\mathcal R_g=R_g I_2$, 
$G_g=I_2$, where $I_2$ is the $2 \times 2$ identity matrix.The energy rate satisfies $\dot H_g=v_g^{\!\top}i_g-R_g\|i_g\|^2$, confirming passivity from effort input $v_g$ to flow output $i_g$.

\subsection{Converter and Filter Subsystem $\Sigma_{\text{conv}}$}

The converter subsystem represents the controlled AC/DC interface, consisting of the filter branch $(R_f,L_f)$ and the averaged converter terminal voltage $e=[e_\alpha,e_\beta]^\top$.
As shown in Fig.~\ref{ac/dc}, the variable $v_{ac}$ denotes the common AC node voltage between the grid-side filter and the converter-side terminals.
The inductor voltage drop across $(R_f,L_f)$ is therefore $(v_{ac}-e)$.

\subsubsection*{1) Physical Model}
Applying Kirchhoff’s voltage law to the filter branch yields
\begin{equation}
L_f \dot i_f = -R_f i_f + v_{ac} - e,
\label{eq:kvl_filter}
\end{equation}
where $i_f=[i_{f\alpha},i_{f\beta}]^\top$ is the filter current.
This equation describes the current dynamics of the converter filter interacting with the AC node.

\subsubsection*{2) Energy Variable and Hamiltonian}
Following the Port-Hamiltonian framework, the magnetic flux linkage of the filter inductor,
\begin{equation}
\phi_f = L_f i_f,
\end{equation}
is selected as the energy (state) variable.
The stored magnetic energy is defined as the Hamiltonian
\begin{equation}
H_p(\phi_f)=\tfrac{1}{2L_f}\|\phi_f\|^2 .
\label{eq:Hp}
\end{equation}
The corresponding co-energy variable (generalized effort) is
\(\nabla_{\phi_f}H_p = i_f = \phi_f/L_f.\)

\subsubsection*{3) Port-Hamiltonian Formulation}
Substituting \(\nabla H_p = i_f\) into \eqref{eq:kvl_filter} gives the state-space relation in PH form:
\begin{equation}
\dot{\phi}_f = -R_f I_2\, \nabla H_p + (v_{ac}-e).
\label{eq:ph_conv_basic}
\end{equation}
Introducing the standard PH matrices
\(
\mathcal J_p=0,\;
\mathcal R_p=R_f I_2,\;
G_p=I_2,
\)
\eqref{eq:ph_conv_basic} can be rewritten compactly as
\begin{equation}
\dot{\phi}_f=(\mathcal J_p-\mathcal R_p)\nabla H_p + G_p(v_{ac}-e).
\label{eq:ph_conv_matrix}
\end{equation}

\subsubsection*{4) DC-Side Port and Power-Preserving Interconnection}
The converter also interfaces with the DC link through an idealized power-preserving map between its AC and DC ports:
\begin{equation}
(v_{ac}-e)^{\!\top}i_f = \eta\,v_{dc} i_{\text{conv}}, \qquad 0<\eta\le1,
\label{eq:power_balance}
\end{equation}
where $i_{\text{conv}}$ is the converter current injected into the DC link and $\eta$ represents conversion efficiency ($\eta=1$ for a lossless converter).
To incorporate this DC port into the PH structure, an additional input matrix $G_{dc}$ is defined such that
\(
G_{dc}=[0,0,1]^\top
\)
and the DC current $i_{dc}$ enters the system as a flow variable.
The complete PH representation of the converter subsystem is then given by
\begin{equation}
\Sigma_{\text{conv}}:
\left\{
\begin{aligned}
\dot{\phi}_f &= (\mathcal{J}_p-\mathcal{R}_p)\nabla H_p \\
&+ G_p(v_{ac}-e) + G_{dc}(-i_{dc})\\
y_p &= 
\begin{bmatrix} i_f \\ v_{dc} \end{bmatrix}
= 
\begin{bmatrix} G_p^{\!\top}\nabla H_p \\ G_{dc}^{\!\top}\nabla H_{dc} \end{bmatrix}
\end{aligned}
\right.
\label{eq:ph_conv_full}
\end{equation}

where $H_p$ and $H_{dc}$ denote the magnetic and electric energy functions, respectively.
The subsystem described by \eqref{eq:ph_conv_full} exchanges power through both AC and DC ports while preserving the overall energy balance dictated by \eqref{eq:power_balance}.

\vspace{-15pt}
\subsection{DC-Link and Load Subsystem $\Sigma_{\text{dc}}$}

The DC-link subsystem consists of the capacitor $C_{dc}$ and the aggregated load representing the data-center IT racks. 
The capacitor acts as an energy storage element that maintains the DC-bus voltage, while the load represents the aggregated downstream converters and servers drawing power from the DC bus.

\subsubsection*{1) Energy Model}
Let the charge on the DC capacitor be $q_{dc}=C_{dc}v_{dc}$. 
The stored electric energy, or Hamiltonian, is
\begin{equation}
H_{dc}(q_{dc})=\tfrac{1}{2C_{dc}}\,q_{dc}^2 .
\label{eq:Hdc}
\end{equation}
The conjugate effort variable is the DC voltage,
\(\nabla_{q_{dc}}H_{dc}=v_{dc}\).
With the converter current $i_{\text{conv}}$ flowing into the capacitor and the load current $i_{\text{load}}$ flowing out, the charge dynamics are
\begin{equation}
\dot q_{dc}=i_{\text{conv}}-i_{\text{load}},
\qquad
y_{dc}=v_{dc}.
\label{eq:dclink_dynamics}
\end{equation}
Equations \eqref{eq:Hdc}–\eqref{eq:dclink_dynamics} define the PH structure of the DC-link subsystem:
\[
\Sigma_{\text{dc}}:\;
\begin{cases}
\dot q_{dc}=G_{\text{conv}}\,i_{\text{conv}}+G_{\text{load}}\,(-i_{\text{load}}),\\[2pt]
y_{dc}=G_{\text{conv}}^{\!\top}\nabla H_{dc}=v_{dc},
\end{cases}
\]
with $G_{\text{conv}}=G_{\text{load}}=1$.

\subsubsection*{2) Non-Passivity of Constant-Power Loads}
In data-center applications, the downstream IT equipment is typically supplied through regulated DC/DC converters operating in constant-power mode.
Such a load can be expressed as
\begin{equation}
i_{\text{load}}=\frac{P_{\text{load}}}{v_{dc}},
\label{eq:CPL}
\end{equation}
where $P_{\text{load}}$ is the demanded active power.
Substituting \eqref{eq:CPL} into \eqref{eq:dclink_dynamics} gives
\begin{equation}
C_{dc}\dot v_{dc}=i_{\text{conv}}-\frac{P_{\text{load}}}{v_{dc}},
\label{eq:CPL_dynamics}
\end{equation}
which yields the instantaneous power balance
\[
\dot H_{dc}=v_{dc}\,i_{\text{conv}}-P_{\text{load}}.
\]
The term $-P_{\text{load}}$ acts as a negative power sink that does not depend on $v_{dc}$ linearly.
Consequently, the incremental relation between $(v_{dc},i_{\text{load}})$ is negative,
\[
\frac{\partial i_{\text{load}}}{\partial v_{dc}}=-\frac{P_{\text{load}}}{v_{dc}^2}<0,
\]
which violates the passivity condition and can lead to oscillatory or unstable DC-link dynamics.
This non-passive behavior of constant-power loads is a critical challenge in tightly regulated DC systems such as data centers.

\subsection{Lossless Interconnection and Overall PH Model}

The three Port-Hamiltonian subsystems $\Sigma_{\text{grid}}$, $\Sigma_{\text{conv}}$, and $\Sigma_{\text{dc}}$ are interconnected through power-preserving Dirac structures that enforce Kirchhoff’s laws at the AC node and the AC/DC interface. 
At the interconnection ports, the voltage and current constraints satisfy
\begin{align}
v_g - v_{ac} &= L_g \dot i_g + R_g i_g, \label{eq:kvl_grid}\\
v_{ac} - e   &= L_f \dot i_f + R_f i_f, \label{eq:kvl_filter2}\\
i_g &= i_f, \label{eq:kcl_ac}\\
(v_{ac}-e)^{\!\top}i_f &= \eta\,v_{dc}\,i_{\text{conv}}, \qquad 0<\eta\le1, \label{eq:powerlink2}
\end{align}
where $\eta$ represents the converter efficiency. 
Equations \eqref{eq:kvl_grid}–\eqref{eq:powerlink2} guarantee that the algebraic sum of the port powers equals zero, thereby ensuring a \emph{lossless interconnection} between the subsystems.

The total stored energy of the composite system is expressed as the sum of all individual Hamiltonians,
\begin{equation}
H_{\text{tot}} = H_g + H_p + H_{dc},
\label{eq:Htot}
\end{equation}
where $H_g$, $H_p$, and $H_{dc}$ denote the magnetic and electric energy functions of the grid, converter, and DC-link subsystems, respectively.
Differentiating \eqref{eq:Htot} along the trajectories of the interconnected system yields the global energy balance
\begin{equation}
\dot H_{\text{tot}} =
v_g^{\!\top}i_g
- R_g\|i_g\|^2
- R_f\|i_f\|^2
- (1-\eta)v_{dc}i_{\text{conv}}
- v_{dc}i_{\text{load}}.
\label{eq:energybalance}
\end{equation}
The first term represents the input power drawn from the grid, and the subsequent terms correspond to dissipative elements associated with the line, filter, and converter losses. 
The last term $-v_{dc}i_{\text{load}}$ accounts for the power extracted by the data-center load.

\subsubsection*{1) Passivity Assessment}
For a conventional resistive load, the system in \eqref{eq:energybalance} is passive from $(v_g,-i_{\text{load}})$ to $(i_g,v_{dc})$.
In data-center systems, however, the downstream regulated converters behave as constant-power loads (CPLs) satisfying
\begin{equation}
i_{\text{load}} = \frac{P_{\text{load}}}{v_{dc}},
\label{eq:CPL2}
\end{equation}
whose incremental slope $\partial i_{\text{load}}/\partial v_{dc}<0$ violates passivity.
This negative incremental resistance causes the DC-link subsystem $\Sigma_{\text{dc}}$ to inject energy rather than dissipate it, potentially destabilizing the entire grid–converter–load system despite the passivity of other branches.

\subsubsection*{2) Control Objective}
The control objective is twofold:  
(i) regulate the DC-bus voltage $v_{dc}$ to its nominal reference $v_{dc}^\star$; and  
(ii) render the DC port of the converter subsystem $\Sigma_{\text{conv}}$ passive so that the total grid--converter--load interconnection preserves a positive energy balance even under non-passive constant-power loads.
The proposed design maintains the PH structure through energy shaping and damping injection.
\vspace{-10pt}
\section{Passivity-Based Controller Design}
The rectifier employs a PH-consistent double-loop structure:
an \emph{outer voltage loop} (energy shaping) and an \emph{inner current loop} (damping injection).
The overall controlled system can be interpreted as the interconnection of the physical converter $\Sigma_{\text{conv}}$ and a PH controller subsystem $\Sigma_{\text{C}}$.

\paragraph*{Controller Subsystem $\Sigma_{\text{C}}$}
The controller stores artificial energy and introduces virtual dissipation through the states 
$\zeta_v$ (voltage loop) and $\zeta_i$ (current loop):
\[
\Sigma_{\text{C}}:
\begin{cases}
\dot\zeta_v = v_{dc}-v_{dc}^\star,\\[2pt]
\dot\zeta_i = i_f - i_f^\star,\\[2pt]
H_{\text{C}}(\zeta_v,\zeta_i)
   = \tfrac{1}{2}a_v\zeta_v^2+\tfrac{1}{2}\zeta_i^\top M_i\zeta_i,\\[2pt]
\tau_{dc} = -k_v(v_{dc}-v_{dc}^\star)-a_v\zeta_v,\\[2pt]
\tau_{ac} = -K_i(i_f-i_f^\star)-M_i\zeta_i,
\end{cases}
\]
where $a_v,M_i\succ0$ and $k_v,K_i>0$ are design constants.
The controller ports $(v_{dc},\tau_{dc})$ and $(i_f,\tau_{ac})$ connect losslessly to the converter's DC and AC ports, respectively.
Its power balance satisfies
\[
\dot H_{\text{C}} + v_{dc}\tau_{dc} + i_f^{\!\top}\tau_{ac}
   = -k_v(v_{dc}-v_{dc}^\star)^2 - K_i\|i_f - i_f^\star\|^2 \le 0,
\]
demonstrating that $\Sigma_{\text{C}}$ is strictly passive and dissipative.
\vspace{-10pt}
\subsection{Energy-Shaping Outer Voltage Loop}

From the DC-link dynamics 
$C_{dc}\dot v_{dc}=i_{\text{conv}}-i_{\text{load}}$, 
an equilibrium is desired at $v_{dc}=v_{dc}^\star$ and $i_{\text{conv}}=i_{\text{load}}$.
Define $e_v=v_{dc}-v_{dc}^\star$.
The outer loop shapes the DC-port energy to decay as
\[
\dot H_{dc}=v_{dc}(i_{\text{conv}}-i_{\text{load}})=-k_v e_v^2,
\]
by commanding
\begin{equation}
i_{\text{conv}}^\star = i_{\text{load}} - \tfrac{k_v}{v_{dc}}(v_{dc}-v_{dc}^\star).
\label{eq:outerloop}
\end{equation}
The term $-k_v(v_{dc}-v_{dc}^\star)^2$ acts as a virtual conductance that cancels the negative incremental resistance of a constant-power load, thereby rendering the DC port passive.

The required AC-side power is 
$p^\star=\eta v_{dc} i_{\text{conv}}^\star$, 
and a dVOC-style current reference in $\alpha\beta$ coordinates is generated as
\begin{equation}
i_f^\star = \frac{p^\star}{\|v_{ac}\|^2}v_{ac}
          + \frac{q^\star}{\|v_{ac}\|^2}Jv_{ac},\qquad
J=\!\begin{bmatrix}0&-1\\1&0\end{bmatrix},
\label{eq:dvoc}
\end{equation}
where $q^\star$ is the desired reactive power (typically zero).

\subsection{Damping-Injection Inner Current Loop}

The filter current dynamics are
\begin{equation}
L_f \dot i_f = -R_f i_f + v_{ac} - e.
\label{eq:ifdyn}
\end{equation}
To track $i_f^\star$ and inject damping, define $e_i=i_f-i_f^\star$ and choose
\begin{equation}
e = v_{ac} - L_f(\dot i_f^\star - K_i e_i) - R_f i_f,
\label{eq:innerlaw}
\end{equation}
with $K_i>0$. 
Substituting \eqref{eq:innerlaw} into \eqref{eq:ifdyn} gives
$L_f\dot e_i=-K_iL_f e_i$, 
an exponentially stable error system that provides positive dissipation 
$-K_i\|e_i\|^2$ at the AC port.
\vspace{-10pt}
\subsection{Closed-Loop Passivity and Grid Interaction}

Connecting $\Sigma_{\text{C}}$ with the PH plant through lossless ports yields the closed-loop storage
\[
H_{\text{cl}} = H_g + H_p + H_{dc} + H_{\text{C}},
\]
whose time derivative is
\begin{align}
\dot H_{\text{cl}}
 &= v_g^{\!\top}i_g
 - (R_g+R_f)\|i_f\|^2
 - (1-\eta)v_{dc}i_{\text{conv}} \notag\\
 &\quad - k_v(v_{dc}-v_{dc}^\star)^2
 - K_i\|e_i\|^2.
\label{eq:Hclfinal}
\end{align}
The term $v_g^{\!\top}i_g$ is the instantaneous power supplied by the grid, 
and all remaining terms are strictly dissipative. 
When $v_g\!\equiv\!0$, the right-hand side of \eqref{eq:Hclfinal} is negative definite, 
guaranteeing asymptotic stability and voltage regulation. 
For time-varying grid voltage amplitude and frequency, 
$v_g(t)$ enters \eqref{eq:Hclfinal} only as an external power input, 
so the closed loop remains \emph{input-strictly passive}: 
the stored energy cannot grow without supplied grid energy.
Hence the converter interacts dissipatively with the grid under any voltage or frequency variation.
\begin{figure}[t]
\centering
\begin{subfigure}[b]{0.85\columnwidth}
    \centering
    \includegraphics[width=\linewidth]{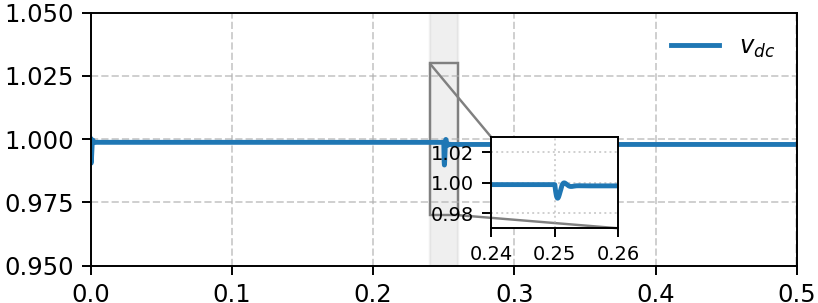}
    \caption{DC load bus voltage (p.u.) with load change.}
\end{subfigure}
\begin{subfigure}[b]{0.85\columnwidth}
    \centering
    \includegraphics[width=\linewidth]{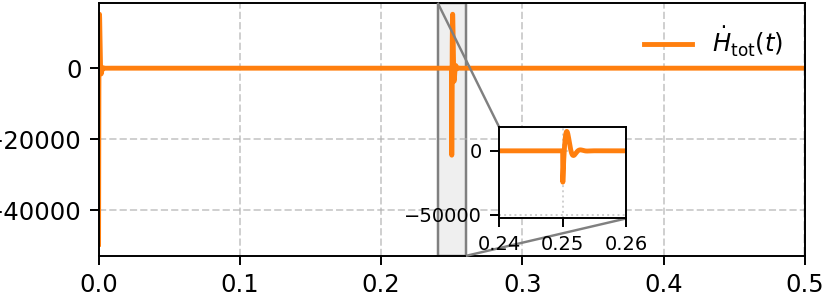}
    \caption{Value of $\dot{H}(t)$.}
\end{subfigure}
\caption{System normal operation performance.}
\label{normal}
\end{figure}

\vspace{-10pt}
\section{Case Study}
The proposed PH based controller is evaluated using a typical data center architecure in Fig.~\ref{ac/dc} and the time-domain simulations conducted using the averaged $\alpha\beta$-frame model described in Section IV. The AC grid is modeled as a 480 $V_{ac}$, 60 Hz source with line impedance, and the DC bus nominal voltage is set to 800 $V_{dc}$. The load represents the aggregated IT racks in a data-center subsystem.
\vspace{-10pt}
\subsection{Normal operation}
This test evaluates the steady-state performance of the proposed controller under nominal grid conditions. The grid voltage is ideal without disturbances, and the primary objective is to verify equilibrium stability and voltage regulation with the DC load  change.

As shown in Fig.~\ref{normal}(a), the DC-bus voltage remains tightly regulated around its nominal value of 1.0~p.u. when the load increases from 1.0~p.u. to 1.5~p.u., exhibiting only a minor undershoot followed by rapid recovery. This demonstrates the controller’s ability to maintain voltage stability and suppress oscillations. Fig.~\ref{normal}(b) presents the corresponding time evolution of $\dot{H_{tot}}$, the derivative of the total stored energy. The short negative excursion in $\dot{H_{tot}}$ indicates that the system dissipates excess energy during the transient before reaching a new equilibrium. Both figures confirm that the converter–load interconnection remains energetically passive and the overall system preserves stability and power balance under normal operation.

 \vspace{-15pt}
 \subsection{Dynamic Performance Under OCP Load Profile} 
\begin{figure}[ht]
\centering 
  \begin{subfigure}{\columnwidth}
   \centering
   {\includegraphics[width=0.8\columnwidth]{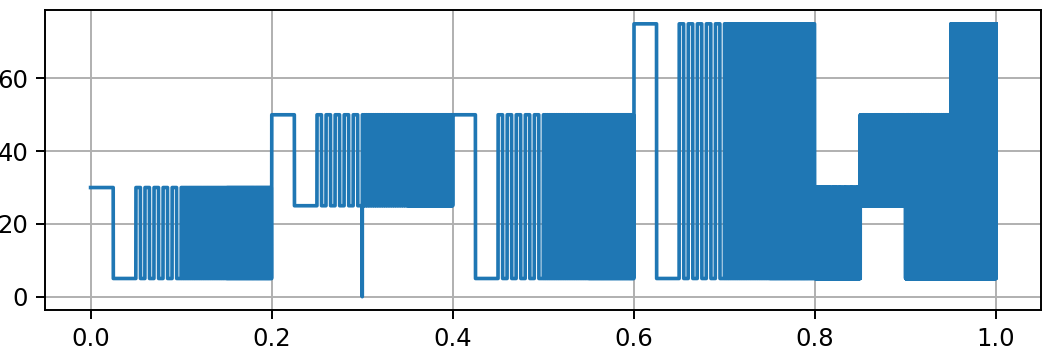}}
  \caption{OCP load profile (kW) }  
  \end{subfigure}
  \begin{subfigure}{\columnwidth}
   \centering
   {\includegraphics[width=0.8\columnwidth]{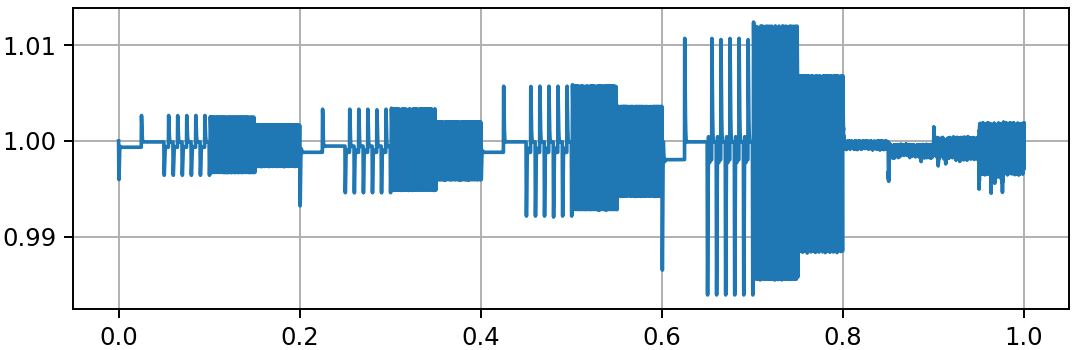}}
  \caption{DC load bus voltage (p.u.) with OCP load profile}  
  \end{subfigure}
\caption{System performance  with OCP load profile\cite{OCP_Diablo400_2025}} 
\label{ocp}
\end{figure}

The applied load follows the measured OCP load profile shown in Fig.~\ref{ocp}(a), which captures the rapidly fluctuating power demand of high-performance computing clusters. As shown in Fig.~\ref{ocp}(b), the DC-bus voltage remains tightly regulated within $1\%$ of its nominal value despite aggressive load transients. The proposed Port-Hamiltonian (PH) controller effectively mitigates the negative incremental resistance introduced by the CPL, providing sufficient virtual damping to suppress oscillations and prevent instability. The small, bounded voltage fluctuations demonstrate that the system preserves energy passivity and dissipative behavior even under highly dynamic load conditions. These results confirm that the proposed passivity-rendering control achieves robust voltage regulation and stable converter–load interaction for data-center–scale variable loads.
\vspace{-15pt}
  \subsection{System Performance with Grid Voltage Sag }
 This scenario evaluates the resilience of the system under grid disturbances. A 20\% grid voltage sag is applied from  t=0.5 s. These conditions emulate typical power quality events in data-center distribution networks.

During the disturbance, the PH-based controller maintains input-strict passivity: the total stored energy temporarily increases due to external excitation but decays immediately after the fault is cleared. The DC voltage remains within ±2\% of its nominal value, demonstrating robust ride-through capability without parameter adaptation.  
 \begin{figure}[ht]
\centering 
  \begin{subfigure}{\columnwidth}
   \centering
   {\includegraphics[width=0.8\columnwidth]{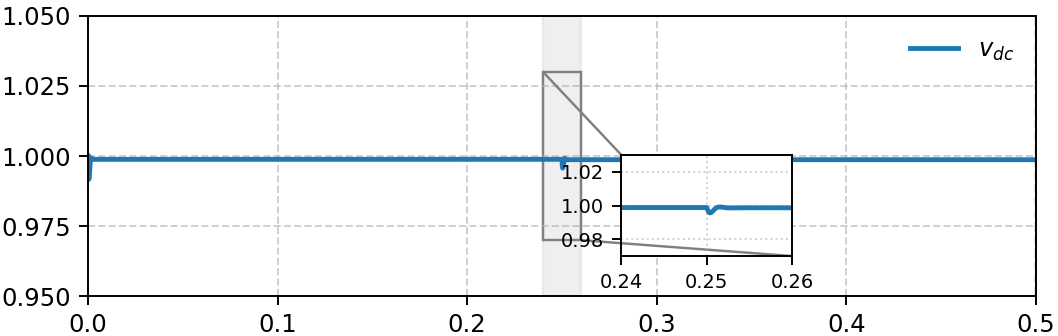}}
  \caption{DC bus voltage (p.u.) with grid voltage sag}  
  \end{subfigure}
  \begin{subfigure}{\columnwidth}
   \centering
   {\includegraphics[width=0.8\columnwidth]{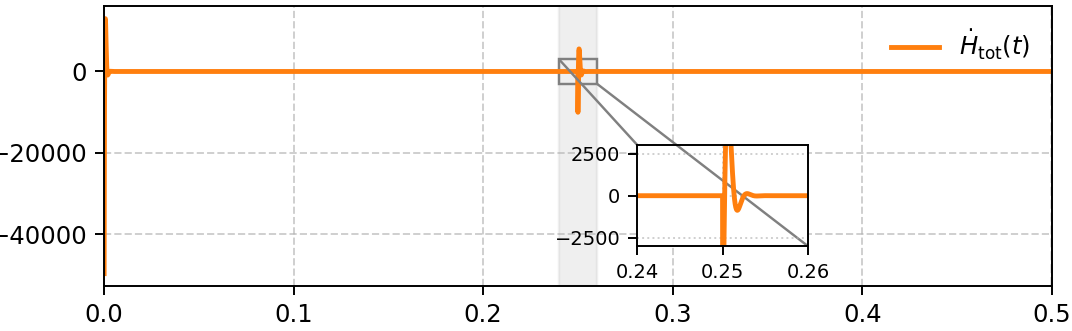}}
  \caption{Value of $\dot{H}(t)$ }  
  \end{subfigure}
\caption{System performance with grid voltage sag}  
\end{figure} 
\vspace{-15pt}
\section{Conclusion}
 This paper presents a Port-Hamiltonian–structured control framework that actively renders grid-tied AC/DC converters passive when supplying constant-power loads in DC data centers.  The framework provides a physics-based, mathematically rigorous foundation for stabilizing non-passive converter–load interactions and achieving resilient grid ride-through. Future work will extend this energy-structured design to large-scale, multi-converter DC architectures, investigate adaptive passivity rendering under parameter uncertainty, and experimentally validate the approach in hardware-in-the-loop and data-center-scale testbeds.
\vspace{-10pt}
\bibliographystyle{ieeetr}
\bibliography{bibliography.bib} 
 
\end{document}